\documentclass[namedreferences]{solarphysics}
\usepackage[optionalrh]{spr-sola-addons} 
\usepackage{epsfig}          
\usepackage{graphicx}        
\usepackage{courier}         
\usepackage{amssymb}        
\usepackage{color}           
\usepackage{url}             
\newcommand{\etal}{{\it et al.}}
\newcommand{\be}{\begin{equation}}
\newcommand{\ee}{\end{equation}}
\newcommand{\beq}{\begin{eqnarray}}
\newcommand{\eeq}{\end{eqnarray}}

\begin{document}

\begin{article}

\begin{opening}

\title{Modification of Proton Velocity Distributions by Alfv\'{e}nic
Turbulence in the Solar Wind}

\author{V.~Pierrard$^{1,2}$ \sep
      Y.~Voitenko$^{1}$ \sep 
   }


\runningauthor{V. Pierrard, Y. Voitenko}
\runningtitle{Proton Velocity Distribution in the Solar
Wind}

\institute{$^1$ Solar-Terrestrial Centre of Excellence, Space Physics, Belgian Institute for Space Aeronomy, Ringlaan 3 av. Circulaire, B-1180 Brussels, Belgium, email: \url{viviane.pierrard@oma.be}
\\
$^{2}$ Georges Lema\^\i tre Centre for Earth and Climate Research (TECLIM), Universit\'e Catholique de Louvain, Place Louis Pasteur 3, bte L4.03.08, 1348 Louvain-la-Neuve, Belgium\\}

\date{Received ; accepted }

\begin{abstract}
In the present paper, the proton velocity distribution function (VDF) in the
solar wind is determined by solving numerically the kinetic evolution
equation. We compare the results obtained when considering the effects of
external forces and Coulomb collisions with those obtained by adding
effects of Alfv\'{e}n wave turbulence. We use Fokker-Planck diffusion terms
due to Alfv\'{e}nic turbulence, which take into account observed turbulence
spectra and kinetic effects of finite proton gyroradius. Assuming a
displaced Maxwellian for the proton VDF at the simulation boundary at 14
solar radii, we show that the turbulence leads to a fast (within several
solar radii) development of the anti-sunward tail in the proton VDF. Our
results provide a natural explanation for the nonthermal tails in the
proton VDFs, which are often observed in-situ in the solar wind beyond 0.3 AU.
\end{abstract}


\end{opening}



\section{Introduction}

The solar wind is a low density plasma in which kinetic processes prevail.
Kinetic models based on the solution of the Fokker-Planck equation have been
developed to study the steady state electron velocity distribution function (VDF) in the corona and at larger
radial distances in the solar wind (SW) \cite{Lies97, Pier99, Lies00, Pier01, Vock03,
Vock09}. The test electrons were submitted to the influence of the external
forces and to Coulomb collisions with background particles \cite{Pier99,
Lies00, Pier01}. Such models were solved numerically and emphasized the
effects of Coulomb collisions compared to the results of purely exospheric
models that considered only the external forces \cite{Maks97, Lamy03}. They
showed that Coulomb collisions have important effects on angular scattering
({\it i.e.} on the pitch angle distribution of the electrons) but do not modify
their average density and mean temperatures radial distributions. Models
including Coulomb collisions gave more realistic VDFs and a reduction of the
temperature anisotropies compared to the purely exospheric approximation
\cite{Lema01}, because the Coulomb collisions isotropize the electron VDF.
Terms due to whistler wave turbulence have also been included in the
electron kinetic equation and its influence on the formation of nonthermal
tails and on the velocity-space diffusion of electrons has been investigated
\cite{PiLa2011}.

In the present paper, we study the evolution of the proton VDF in the solar
wind. First, we use a kinetic approach similar to the mentioned above
papers, which accounts for the effects on the protons of the external forces
and Coulomb collisions. After that we include also the wave effects on the
proton VDF. Recent in-situ measurements have revealed that the wave activity
at the proton kinetic scales is dominated by Alfv\'{e}n waves rather than
whistlers \cite{He11, He12, Podesta11}. Among Alfv\'{e}n waves, the most
power, about 80\%, is in highly oblique kinetic Alfv\'{e}n waves (KAWs), and
the rest 20\% is in quasi-parallel ion-cyclotron Alfv\'{e}n mode \cite{He11}%
. Both the MHD-scale and kinetic-scale Alfv\'{e}n waves posses power law
spectral distributions and exhibit turbulence properties. We therefore use
in present simulations a 1D-2D (1D spatial - 2D velocity-space)
Fokker-Planck diffusion term due to Alfv\'{e}nic turbulence derived by
Voitenko and Pierrard (2012).

Voitenko and Goossens (2006) suggested that the proton trapping in the
parallel KAW potential can be responsible for the proton beams observed in
the solar wind. Later on, at the Solar Wind 12 in June 2009, Pierrard and
Voitenko have presented analytical estimations and numerical example,
published in Pierrard and Voitenko (2010). Numerical simulations by Osmane
\textit{et al.} (2010) and Li \textit{et al.} (2010) have demonstrated the
proton beam formation by a coherent monochromatic KAW. Our present approach
is based not on monochromatic KAWs but on a wide spectrum of KAWs with
overlapping potentials, which results in the velocity-space proton diffusion
rather than the proton trapping as in the mentioned above papers.

\section{Description of the Model}

The kinetic transport equation for the evolution of the velocity
distribution function $f(\mathbf{r,v},t)$ of the protons in the SW is given
by:
\begin{equation}
\frac{\partial f(\mathbf{r,v},t)}{\partial t}+(\mathbf{v.\nabla _{r}})f(%
\mathbf{r,v},t)+(\mathbf{a.\nabla _{v}})f(\mathbf{r,v},t)=\left( \frac{df}{dt%
}\right)_{\rm C}+\left( \frac{df}{dt}\right)_{\rm A}  \label{e1}
\end{equation}%
where $\mathbf{r}$ and $\mathbf{v}$ are respectively the position and
velocity vectors of the particles, $\mathbf{a}$ is the acceleration due to
external forces and $t$ is the time. We are interested by the steady state
solution of this equation. The right-hand-side represents the interactions
between the particles. In the present paper, we consider the Coulomb
collisions with the Fokker-Planck term $\left( \frac{df}{dt}\right) _{\rm C}$
and wave-particle interactions with kinetic Alfv\'{e}n waves $\left( \frac{df%
}{dt}\right)_{\rm A}$.

In the case of the SW, the forces are the electric force $Ze\mathbf{E}$, the
gravitational force $m\mathbf{g}$ and the Lorentz force resulting from the
magnetic field distribution. The acceleration term for the protons is thus
given by:
\begin{equation}
\mathbf{a}=\left( \frac{Ze\mathbf{E}}{m}+\mathbf{g}\right) +\frac{Ze}{m}(%
\mathbf{v}\times \mathbf{B})=\mathbf{a}_{r}(r)+\mathbf{a}_{\rm L}  \label{e2}
\end{equation}%
where $Ze$ is the electric charge of the protons ($Z=1$), $m$ their mass, $%
\mathbf{E}$ the electric field, $\mathbf{g}$ the gravitational acceleration,
$\mathbf{B}$ the magnetic field (assumed to decrease as $r^{-2}$), $\mathbf{%
a}_{r}(r)$ is the radial component of the acceleration due to the electric
and gravitational forces that are vertical and $\mathbf{a}_{\rm L}$ is the non
radial term due to the Lorentz force.

The simplest solar wind models consider only the effects of these external
forces. In such models called exospheric, the interaction terms in Equation (\ref%
{e1}) are neglected so that an analytic solution of the equation can be
obtained.

\begin{figure}[tbp]
\caption{Profiles of (a) number density, (b) electric potential, (c) flux, (d) bulk speed,
(e) proton temperature $T_{\rm p}$ (upper solid line: $T_{\parallel}$, bottom solid line:
$T_{\perp}$, dotted line: average $T$), (f) electron temperature ({\it idem}),
(g) proton potential, and (h) heat flux (upper and lower curves for electrons and
protons) from $r_0$= 1.5 $R_{\rm s}$ up to 200 $R_{\rm s}$ obtained with a kinetic exospheric
model of the solar wind with $T_{\rm e}=1.5 \times 10^6$ K and $T_{\rm p}=1.5 \times 10^6$
K for a Maxwellian proton VDF and a Kappa electron VDF with $\protect\kappa$%
=3.}
\label{bc2}
\begin{center}
\includegraphics[width=11.5cm]{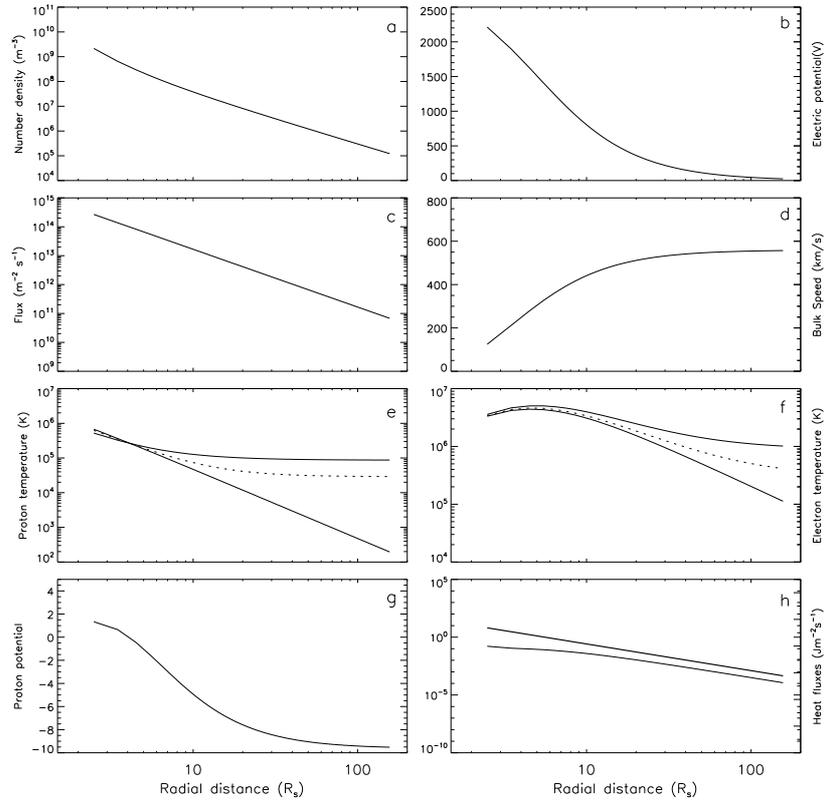}
\end{center}
\end{figure}

Figure 1 shows the profiles of the VDF moments obtained with the exospheric
solar wind model developed by Pierrard and Lemaire (1996) and improved by
Maksimovic \textit{et al.} \ (1997) and Lamy \textit{et al.} (2003). In such
models, truncated Maxwellian and Kappa VDF are assumed respectively for the
protons and the electrons with given density and temperatures at the exobase
level (here chosen to be at 1.5 $R_{\rm s}$ (solar radius)). The VDF is obtained at
larger radial distance by solving the evolution equation, {\it i.e.} Equation (\ref{e1}%
) without interaction right-hand-side terms.

It can be seen (Figure 1d) that the particles are accelerated at low radial
distances due to the effects of the external forces. It is mainly from 1 to
20 $R_{\rm s}$ that the bulk velocity of the particle increases and then it becomes
almost constant at higher radial distances. This result is similar to that
obtained in MHD solar wind models where the interactions between the
particles are taken into account \cite{Park1958}.

The exospheric model emphasizes that it is the electrostatic potential (see
Figure 1b) that increases the velocity of the particles. This potential
ensures that the proton flux is everywhere equal to that of the electrons so
that no net current is transported by the solar wind.

In this model, the acceleration is reinforced by the presence of
suprathermal electrons simulated by a Kappa VDF. They increase the
electrostatic potential and thus accelerate the wind. On the contrary, the
presence of suprathermal protons has almost no influence on the
electrostatic potential \cite{Maks97}. That is why a Maxwellian is assumed
at the reference level for the proton VDF at 1.5 $R_{\rm s}$. The averaged values of
the different moments (profiles of number density, bulk velocity,
temperatures) are in good agreement with the solar wind observations, except
the temperature anisotropies that are too high in the model and would be
expected to produce plasma instabilities in the system. Such high
anisotropies are mainly due to the assumption of magnetic moment
conservation for each particle species that is not realistic, even in low
density plasmas.

\section{Numerical Model}

More sophisticated solar wind models including interaction terms (the right-hand-side of Equation (1)) need to be solved numerically. We develop in the present
paper such a numerical model for the protons, based on similar assumptions
as those used for the model of SW electrons \cite{Pier99}.

We simplify the equation by using only three coordinates: the radial
distance $r$, the velocity $v$ and $\eta =\cos \theta =\hat{\mathbf{B}}%
_{0}\cdot \hat{\mathbf{k}}$ where $\theta $ is the pitch angle between the
velocity vector and the magnetic field direction assumed to be radial. The
left-hand-side of Equation (1) becomes:
\begin{equation}
{\rm D}f=\frac{\partial f}{\partial t}+v\eta \frac{\partial f}{\partial r}%
+a_{r}(r)\left( \eta \frac{\partial f}{\partial v}+\frac{(1-\eta ^{2})}{v}%
\frac{\partial f}{\partial \eta }\right) +\frac{v}{r}(1-\eta ^{2})\frac{%
\partial f}{\partial \eta }.  \label{e3}
\end{equation}

As a first study, we will consider that the term $(df/dt)_{\rm C}$ on the right-hand-side of Equation (\ref{e1}) represents the effects of the Coulomb
collisions {\cite{Pier01}. Then, in a next section of this paper, we adopt
in addition the kinetic Alfv\'{e}n wave term $(df/dt)_{\rm A}$. }

\section{Spectral Method of Solution}

To obtain steady state solutions for the proton VDF, we use a specialized
spectral method similar as that described in \cite{Pier11} for the
resolution of the Fokker-Planck equation (FPE).

The dimensionless velocity is defined by
\begin{equation}
x=\sqrt{\frac{m_{\rm p}}{2k_{\rm B}T_{\rm p}(r)}} v=\frac{v}{v_{\rm th}(r)},  \label{e10}
\end{equation}
where $v_{\rm th}$ is the proton thermal speed and $k_{\rm B}$ is the Boltzmann
constant.

The solution is expanded in terms of orthogonal polynomials:
\begin{equation}
f(z,x,\eta )=\exp (-x^{2})\times \left(
\sum_{l=0}^{J-1}\sum_{s=0}^{N-1}\sum_{m=0}^{M-1}a_{lsm}P_{l}(\eta
)S_{s}(x)L_{m}(z)\right) .  \nonumber  \label{dvlpmt2}
\end{equation}%
where $J$, $N$, and $M$ are integers whose value is adjusted to obtain the
required numerical precision for the solution. We use Legendre polynomials $%
P_{l}(\eta )$ with respect to $\eta =\cos \theta $, speed polynomials $%
S_{s}(x)$ with respect to the normalized velocity $x$ and displaced Legendre
polynomials on the interval $[0,c]$ with respect to the dimensionless
altitude $z$. The $a_{lsm}$ coefficients are the coefficients of the
expansion to be determined.

The boundary condition determines the value of some coefficients $a_{lsm}$
at the reference level. The equation is solved numerically to find the VDF $%
f(z,x,\eta )$ at the other radial distances. We choose 10 polynomials for
each variable so that the results have a precision better than 10 \% with
reasonably short CPU times. We have checked that the results do not change
significantly when $J$, $N$ or $M$ is increased.

The advantage of this method is that in a discrete ordinate basis, the
derivatives of any continuous function $f(y)$ can be approximated by the
following expansion:
\begin{eqnarray}
\left(\frac{\partial f}{\partial y}\right)_{y=y_i}\approx%
\sum_{j=0}^{N-1}D_{ij}f(y_j)  \label{derivee}
\end{eqnarray}
where $D_{ij}$ are the matrix elements of the derivative operator in the
polynomial basis.

The integrals to calculate the moments of the protons are also easily
performed by numerical quadrature taking into account the polynomial
expansion \cite{Pier97}. The moments are calculated from the VDF found by
solving the FPE. The density is given by:
\begin{equation}
n(r)=\int f(r,v,\eta )\ d\mathbf{v}.  \label{dens}
\end{equation}%
The bulk velocity is calculated by:
\begin{equation}
\mathbf{u}(r)=\left( \int f(r,v,\eta )\mathbf{v}d\mathbf{v}\right) /n.
\label{bulk}
\end{equation}%
The temperature corresponds to a second order moment of the VDF:
\begin{equation}
T(r)=\frac{1}{3}(T_{\parallel }(r)+2T_{\perp }(r))  \label{temp}
\end{equation}%
where
\begin{eqnarray}
T_{\parallel }(r) &=&\frac{m\int (v_{\parallel }-u)^{2}f(r,v,\eta )\ d%
\mathbf{v}}{k_{\rm B}n(r)}  \label{tpar} \\
T_{\perp }(r) &=&\frac{m\int v_{\perp }^{2}f(r,v,\eta )\ d\mathbf{v}}{%
2k_{\rm B}n(r)}.  \label{tper}
\end{eqnarray}

\section{The Effects of Coulomb Collisions}

The effects of the binary Coulomb collisions are taken into account by using
as the right-hand-side term in Equation (\ref{e1}) the Fokker-Planck collision
operator appropriate when large-angle deflections can be neglected (Hinton, 1983):
\begin{equation}
\left( \frac{d f}{d t} \right)_{\rm C}= -\frac{\partial }{\partial \mathbf{v}}
\cdot \left[ \mathbf{A} f(\mathbf{r,v},t) -\frac{1}{2}\frac{\partial}{%
\partial \mathbf{v}} \cdot \left( \mathbf{D}f(\mathbf{r,v},t)\right) \right]
\label{FP}
\end{equation}
where $\mathbf{A}$ is the dynamic friction vector%
\begin{eqnarray}
\mathbf{A}&=&-4\pi \frac{Z_{\alpha}^2 Z_{\beta}^2 e^4 \ln \Lambda}{%
m_{\alpha}^2}\left( 1+\frac{m_{\alpha}}{m_{\beta}} \right)  \nonumber \\
&\times &\int d\mathbf{v^{\prime}} f_{\beta}(\mathbf{v^{\prime}})\frac{(%
\mathbf{v}-\mathbf{v^{\prime}})}{(v-v^{\prime})^3}  \label{friction}
\end{eqnarray}
and $\mathbf{D}$ is the velocity diffusion tensor
\begin{eqnarray}
\mathbf{D} &=&4\pi \frac{Z_{\alpha}^2 Z_{\beta}^2 e^4 \ln \Lambda} {%
m_{\alpha}^2}  \label{diffus} \\
&\times &\int d\mathbf{v^{\prime}} f_{\beta}(\mathbf{v^{\prime}})\left(
\frac{\mathbf{I}}{v-v^{\prime}}-\frac{(\mathbf{v}-\mathbf{v^{\prime}})(%
\mathbf{v}-\mathbf{v^{\prime}})}{(v-v^{\prime})^3} \right).  \nonumber
\end{eqnarray}

In Equations (\ref{friction}) and (\ref{diffus}), $\ln \Lambda$ is the usual Coulomb
logarithm containing the Debye screening effect: $\ln \Lambda \approx 24$. The
index $\alpha$ corresponds to the test protons and $\beta$ to the background
particles, $m$ the mass and $Ze$ the charge of the particles, $%
f_{\beta}(v^{\prime })$ is the velocity distribution function of the
background particles assumed to be a displaced Maxwellian distribution since
this VDF is solution of the Fokker-Planck operator.

We assume boundary conditions at a reference level here chosen to be 14 $R_{\rm s}$.
It could be chosen at lower radial distances as well, but this distance
corresponds to the region where kinetic Alfv\'{e}n turbulence is supposed to
have effects, so this altitude is especially appropriate for comparisons
when this term included (see the next section). We assume an isotropic displaced
Maxwellian for the protons VDF at this reference distance since this VDF is
solution of the Fokker-Planck Coulomb collision term. The bulk velocity at
this radial distance is chosen to be 470 km s$^{-1}$, as obtained with the
exospheric model.

Figure 2 illustrates the VDFs (normalized to the thermal velocity and
centered on the averaged bulk velocity) obtained at different radial
distances by solving the FPE including the Coulomb collision term. The VDF
is always presented in the reference frame centered at its bulk velocity.
The dotted circle corresponds to $v/v_{\rm th}=\sqrt{3/2}$.

The VDF remains close to a displaced Maxwellian at higher radial distances
with a bulk velocity slightly increasing as illustrated on Figure 3b. Figure 3a illustrates the profile of the decreasing normalized
number density and Figure 3c the decreasing temperature.
Compared to the exospheric model, the effects of Coulomb collisions
(obtained with the inclusion of the Fokker-Planck collision term and with
the isotropic boundary condition) are mainly to reduce the temperature
anisotropy of the particles.

The results depend on the boundary condition at the reference level. Since
we start with an isotropic displaced Maxwellian at the reference level
chosen to be 14 $R_{\rm s}$, the VDF remains close to a displaced Maxwellian at
higher radial distances even if the Coulomb collisions become negligible at
large radial distances. The bulk velocity increases with the radial distance
due to the acceleration term and the Lorentz force slightly alignes the VDF
to the magnetic field direction with the increasing radial distances.

\begin{figure}[tbp]
\begin{center}
\includegraphics[width=11.5cm]{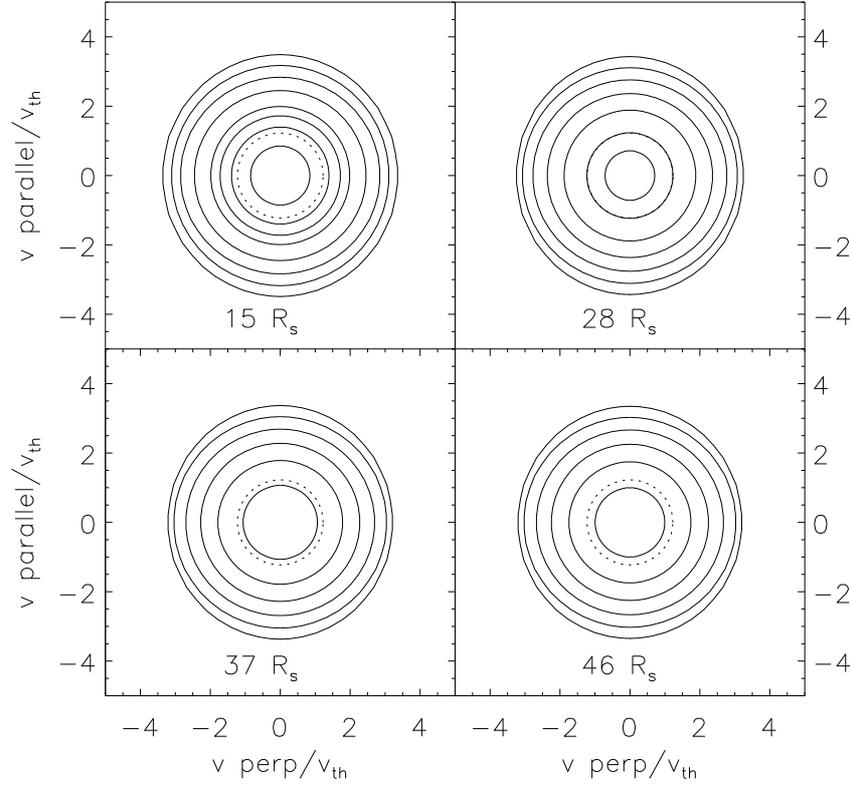} 
\end{center}
\caption{VDFs obtained at different radial distances assuming a displaced
Maxwellian as boundary condition at 14 $R_{\rm s}$ by solving the FPE including the
Coulomb collision term.}
\label{bc2f}
\end{figure}

\begin{figure}[tbp]
\begin{center}
\includegraphics[width=11.5cm]{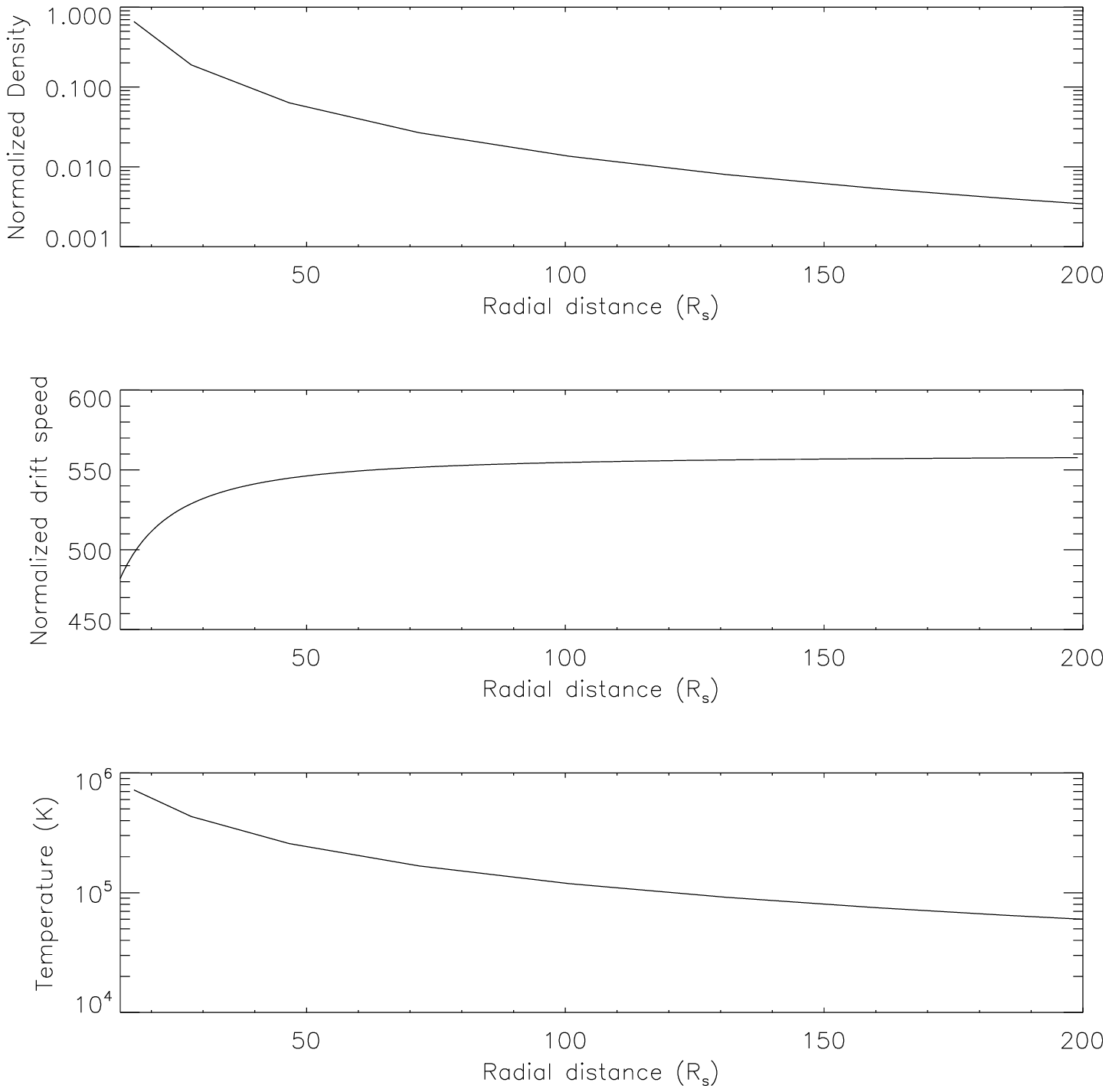}
\end{center}
\caption{Radial profiles of number density, bulk velocity, and proton temperature profiles
obtained by solving the FPE including the Coulomb collisions assuming a
displaced Maxwellian as boundary condition at 14 $R_{\rm s}$.}
\label{mom215}
\end{figure}

\section{Fokker-Planck Diffusion Terms due to Alfv\'{e}nic Turbulence}

In this section we include diffusion terms due to Alfv\'{e}nic turbulence.
The Alfv\'{e}\-nic turbulence evolves towards small scales across the
background magnetic field and at proton kinetic scales transforms into the
turbulence of KAWs. The preliminary principles of
KAWs on the formation of the proton beams have been presented by
\inlinecite{PiVo10}. The focus in this previous paper was on the proton
trapping and acceleration in the electric potential carried by intermittent
isolated KAWs. Here we use a model Alfv\'{e}nic spectrum that corresponds to
the magnetic power spectra measured in the solar wind.

The Alfv\'{e}nic turbulence and related terms describing the velocity-space
proton diffusion in the solar wind conditions have been presented in detail
by Voitenko and Pierrard (2012). We use here the final expressions from this
paper, which are in terms of quantities measured in-situ in the solar wind.
The Fokker-Planck term due to Alfv\'{e}nic turbulence, which is essentially
the KAW turbulence at small scales, is given by

\begin{eqnarray*}
\left( \frac{\partial f_{\rm p}}{\partial t}\right) _{\mathrm{A}} &=&\left( \eta
\frac{\partial }{\partial v}+\frac{1-\eta ^{2}}{v}\frac{\partial }{\partial
\eta }\right) D_{\mathrm{A}}\left( \eta \frac{\partial }{\partial v}+\frac{%
1-\eta ^{2}}{v}\frac{\partial }{\partial \eta }\right) f \\
&=&\eta ^{2}\frac{\partial }{\partial v}D_{\mathrm{A}}\frac{\partial }{%
\partial v}f+\eta \frac{\partial }{\partial v}\frac{1-\eta ^{2}}{v}D_{%
\mathrm{A}}\frac{\partial }{\partial \eta }f \\
&&+\frac{1-\eta ^{2}}{v}\frac{\partial }{\partial \eta }\eta D_{\mathrm{A}}%
\frac{\partial }{\partial v}f+\frac{1-\eta ^{2}}{v}\frac{\partial }{\partial
\eta }\frac{1-\eta ^{2}}{v}D_{\mathrm{A}}\frac{\partial }{\partial \eta }f.
\end{eqnarray*}%
The KAW diffusion coefficient
\[
D_{\mathrm{A}}=\alpha \pi ^{2}\Omega _{\rm p}V_{\rm S}V_{\rm A}\frac{\left( \frac{\left(
\eta v/V_{\rm A}\right) ^{2}-1}{2}\right) ^{0.2-0.55\left( \left( \eta
v/V_{\rm A}\right) ^{2}-1\right) /\left( \left( \eta v/V_{\rm A}\right)
^{2}+1\right) }}{\left( \left( \eta v/V_{\rm A}\right) ^{2}+1\right) \left( \eta
v/V_{\rm A}\right) }\frac{\left\vert B_{\mathbf{1}}\right\vert ^{2}}{B_{0}^{2}}
\]%
for $1\leq \eta v/V_{\rm A}\leq \sqrt{1+2\mu_{\mathrm{c}}^{2}}$, and $D_{%
\mathrm{A}}=0$ otherwise, where $\mu =k_{\perp }\rho _{\rm p}$ is the
perpendicular wavenumber $k_{\perp }$ normalized by the proton gyroradius $%
\rho _{\rm p}$. Other notations are as follows: $\Omega _{\rm p}$ is the proton
cyclotron frequency. $V_{\rm A}$ is the Alfv\'{e}n velocity:

\begin{eqnarray}
V_{\rm A}(\mathrm{cm~s}^{-1})=2.18\times10^{11}\frac{B(\mathrm{gauss})}{\sqrt{n(\mathrm{%
cm^{-3}})}}
\end{eqnarray}
$V_{\rm S}=\sqrt{k_{\rm B} T_{\rm e \parallel}/m_{\rm p}}$ is the ion-acoustic velocity, and the
anisotropy factor $\alpha \approx 0.3$. All the plasma parameters (background
number density $n_{0}$, parallel electron temperature $T_{\rm e \parallel }$, and
background magnetic field $B_{0}$) are functions of $r$.

The above expression for $D_{\mathrm{A}}$ is obtained from Equation (7) by
Voitenko and Pierrard (2012) using a power-law turbulence spectrum $%
\left\vert B_{\mu \mathrm{SC}}\right\vert ^{2}\propto \left\vert B_{\mathbf{1%
}}\right\vert ^{2}\mu ^{-\delta _{\perp }}$ with the following model for the
variable index: $\delta _{\perp }=$ $1.6+1.1\mu ^{2}\left( 1+\mu ^{2}\right)
^{-1}$. This $\delta _{\perp }$ has the low- and high-$\mu $ asymptotes $1.6$
and $2.7$, as is measured in the solar wind (Alexandrova \textit{et al.},
2009). We also averaged $D_{\mathrm{A}}$ in the cross-field plane and put $%
T_{\rm e}/T_{\rm p}=1$ in the KAW dispersion.

A low-beta two-fluid dispersion equation is adopted for KAWs,
\begin{eqnarray}
\frac{\omega }{k_z V_{\rm A}}=\sqrt{1+\left( 1+\frac{T_{\rm e}z}{T_{\rm p \perp }}%
\right) \mu ^{2}},
\end{eqnarray}
which is a good approximation for the kinetic dispersion by Hasegawa and
Chen (1976) where a possible temperature anisotropy is taken into account. A
comparison among several KAW models is given by Voitenko and Pierrard (2012).%

\section{Formulation of the Boundary-value Problem at the Solar Wind Base}

We are interested in the quasi-stationary boundary-value problem, where the
MHD Alfv\'{e}n waves (AWs) are injected along $\mathbf{B}_{0}$ at $r=r_{0}=14\
 R_{\rm S}$. We assume $B(200\ R_{\rm s})=4$ nT $= 4 \times 10^{-5}$ gauss and decreasing as $%
r^{-2}$ to determine $B$ at 14 $R_{\rm s}$. Then $d/dt$ means $v_{\rm z}\partial
/\partial r$ in the Fokker-Planck Equation (\ref{e1}). The nonlinear
interaction of MHD AWs create MHD turbulent cascade towards smaller scales
and turbulent spectra. It is found that the turbulent cascades are
anisotropic and proceed mainly towards small cross-field wavelengths, where
the waves are essentially kinetic Alfv\'{e}n waves undergoing a
wave-particle Cherenkov interaction.

As the plasma parameters, magnetic field $B_{0}$, and Alfv\'{e}n velocity $%
V_{\rm A}$ vary with distance, we expect that the upward propagating KAW
turbulence sweeps the anti-sunward part of the particle distribution
functions making them flatter or even producing a separate beam. Such
distortions of the velocity distributions are often measured at the
distances $\gtrsim 0.3$ AU \cite{Mars82}, which can be signatures of the
velocity diffusion due to the KAW turbulence.

We assume that, contrary to the distant solar wind, the Alfv\'{e}nic
turbulence in the vicinity of the solar wind base does not fill all the
wavenumber range but evolves rapidly. We will model this by introducing a
distance-dependent high-wavenumber cutoff $\mu _{\mathrm{c}}$ for the
spectrum of such "young", not fully developed turbulence. The turbulent
spectrum in the "young" turbulence does not spread beyond $\mu =\mu _{%
\mathrm{c}}$, but $\mu _{\mathrm{c}}$ itself increases with heliocentric
distance. Because of the absence of the observational data on the cutoff $%
\mu _{\mathrm{c}}$ within 0.3 AU, we model it as
\[
\mu _{\mathrm{c}}=\mu _{\mathrm{c}}\left( r\right) =\mu _{\mathrm{c0}}\left(
\frac{r}{r_{0}}\right) ^{3},
\]%
where $\mu _{\mathrm{c0}}$ is the cutoff value at the solar wind base $%
r=r_{0}$. We assume $\mu _{\mathrm{c0}}=0.1$ at the boundary, such that the
turbulent spectrum does not reach the proton kinetic length scales $\approx
\rho_{rm p}$. The proton gyro-scale $\mu _{\mathrm{c}}\approx 1$ is reached by
the turbulent cascade at the distances $\approx 2r_{0}$.

In the near-Earth solar wind at $R=1$ AU the normalized spectral density $%
\left\vert B_{\mathbf{1}}\right\vert ^{2}/B_{0}^{2}\approx 10^{-3}$
(Alexandrova \textit{et al.}, 2010). Since the radial dependence of $%
\left\vert B_{\mathbf{1}}\right\vert ^{2}/B_{0}^{2}$ is not known, we apply
a self-similar projection from 1 AU to $r_{0}$ = 14 $R_{\rm S}$ by keeping
constant the ratio of the total turbulence energy to the spectral density at
$\mu =1$. Then we use a model profile
\begin{equation}
\frac{\left\vert B_{\mathbf{1}}\right\vert ^{2}}{B_{0}^{2}}=0.05\left[
\left( \frac{r}{r_{0}}\right) ^{-0.1}-0.68\right] ,  \label{B1}
\end{equation}%
describing a relatively slow decrease of $\left\vert B_{\mathbf{1}%
}\right\vert ^{2}/B_{0}^{2}$ with $r$ starting from $\left\vert B_{\mathbf{1}%
}\right\vert ^{2}/B_{0}^{2}$ $\approx $ $1.6\times 10^{-2}$ at $r=r_{0}=14\ R
_{\rm S}$, to $\left\vert B_{\mathbf{1}}\right\vert ^{2}/B_{0}^{2}\approx 4\times
10^{-3}$ at 1 AU. Alexandrova \textit{et al.} (2010) reported values $%
\gtrsim 10^{-3}$ at 1 AU. This profile fits the turbulence energy in the
time domains from 3 to 25 minutes, obtained by Bavassano \textit{et al.}
(1982) from Helios measurements at three radial distances between 0.3 and 1
AU. Other attempts to model the development of solar wind turbulence have
adopted other turbulence profiles \cite{Cran03, Cran07}, which imply
qualitatively similar turbulence behavior above 15 $R_{\rm s}$.

The spectral power density of magnetic fluctuations at any scale $%
\mu $ is $W_{\mu }=B_{\mu }^{2}/\mu $, where $B_{\mu }$ is the magnetic
amplitude at that scale. The total energy of the turbulence is given by the
integral over all $\mu $ from the lowest cutoff at electron scales to the
largest (driving) scale at the break in the spectral index between -1 and -5/3.
Most of this energy is carried by the large-scale MHD waves that compose the
MHD inertial range. Formally the energy in the steep kinetic spectra of KAWs
immediately available for particles is much less, of the order of $q^{-2/3}$
(here $q=k_{\perp \rm s}/k_{\perp 1}$ is the driving-to-kinetic scale ratio, or
\textquotedblleft width\textquotedblright\ of the inertial range). The wave
propagation angle at proton kinetic scales is close to $\pi /2$. Our $B_{1}$
is an average fluctuating amplitude at a particular break scale $\mu =\mu
_{1}$ (we assumed $\mu _{1}$=1 for simplicity ). It is therefore much
smaller than the total energy of Alfv\'{e}n waves at all scales. 

\section{Modification of Proton VDF by the Alfv\'{e}nic Turbulence}

It seems that the observed turbulence spectra do not evolve significantly
beyond 0.3 AU, which means they are formed much closer to the Sun. Since the
kinetic-scale Alfv\'{e}nic turbulence and its effects on protons are our
primary interest, it is convenient to set the simulation boundary at the
radial distance where the turbulence cascade approaches kinetic scales. As a
reference level we chose here 14 $R_{\rm s}$ where we allow the turbulence cascade to
spread in the high-wavenumber range. Again, the solar wind is already formed
in this region and the turbulence terms can be included in consideration
without interfering much with the SW acceleration processes. We assume a
displaced Maxwellian for the protons VDF at this reference altitude since
this VDF is solution of the Fokker-Planck Coulomb collision term. The bulk
velocity at this radial distance is chosen to be 470 km s$^{-1}$. This choice is conventional and the boundary
could be at lower or larger radial distances, which however could not change
simulation results except for adjusting them to a new distance.

Figures 4-6 present results of the simulations including the turbulence terms
for the radial distances starting from the simulation boundary at 14 $R_{\rm s}$. At
the boundary, the same displaced Maxwellian for protons was used as in the
previous simulations. The results presented in Figure 4 show the proton VDF
evolution with the radial distance. As compared to the previous simulation
without turbulence, where the proton VDF remains a displaced Maxwellian at
all distances, the presence of the Alfv\'{e}nic turbulence in the solar wind
leads to a fast development of nonthermal tails in the proton VDFs. The
tails are already noticeable at the distances about one solar radius from
the boundary, and they increase fast with the radial distance and become
pronounced after 2-3 solar radii.

The radial behavior of the moments of the proton VDF are shown in 
the top panels of Figure 5. As is seen from Figure 5, the profiles of the bulk
velocity and number density are also modified by the turbulence. The bulk
velocity is increased in the regions where the tails are formed, and it is
stabilized at higher altitudes. The number density profile decreases slower
than without turbulence. The number density profile decreases less fast when
the KAW amplitude $B_{1}/B_{0}$ is chosen to be smaller. The bottom
panels of Figure 5 show the profiles of the Alfv\'{e}n velocity and of the
plasma beta function:
\begin{eqnarray}
\beta =\left( 1+\frac{T_{\rm e\parallel }}{T_{\rm p \perp }}\right) \frac{V_{\rm Tp}^{2}}{%
V_{\rm A}^{2}}
\end{eqnarray}
where $V_{\rm Tp}=\sqrt{k_{\rm B}T_{\rm p}/m_{\rm p}}$. The most favorable
conditions for the tail generation occur in the regions where the proton
thermal and Alfv\'{e}n velocities are about the same, {\it i.e.} $\beta \approx 1$.

\begin{figure}[tbp]
\begin{center}
\includegraphics[width=11.5cm]{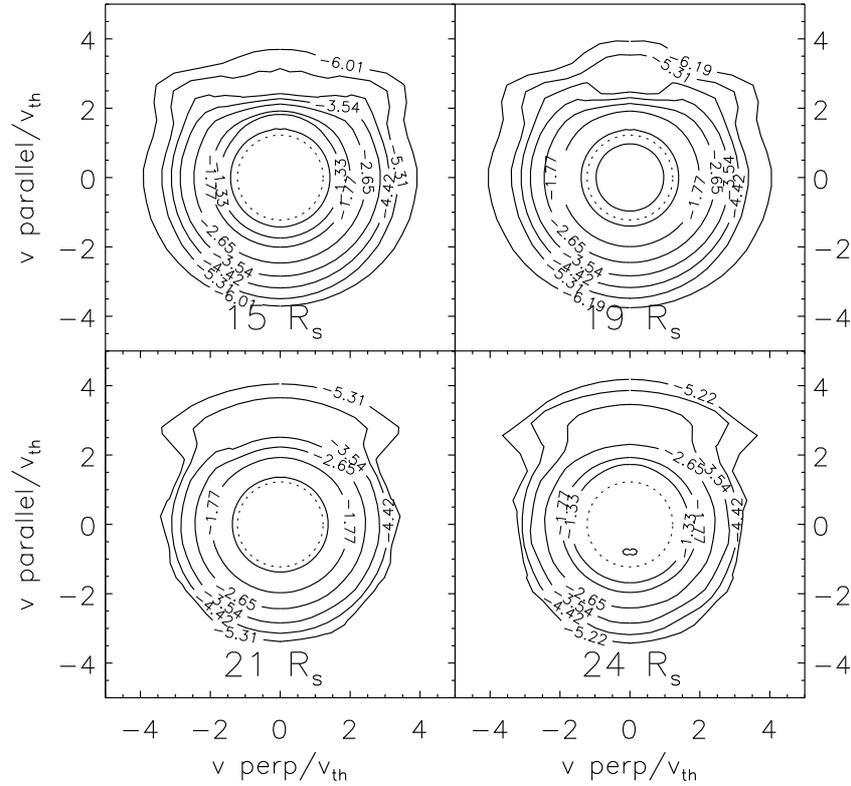}
\end{center}
\caption{Isocontours of the proton VDF obtained at different radial
distances assuming a displaced Maxwellian as boundary condition at 14 $R_{\rm s}$.
The evolution equation was solved numerically and included the Fokker-Planck
diffusion terms due to turbulence.}
\label{bc2f}
\end{figure}

\begin{figure}[tbp]
\begin{center}
\includegraphics[width=11.5cm]{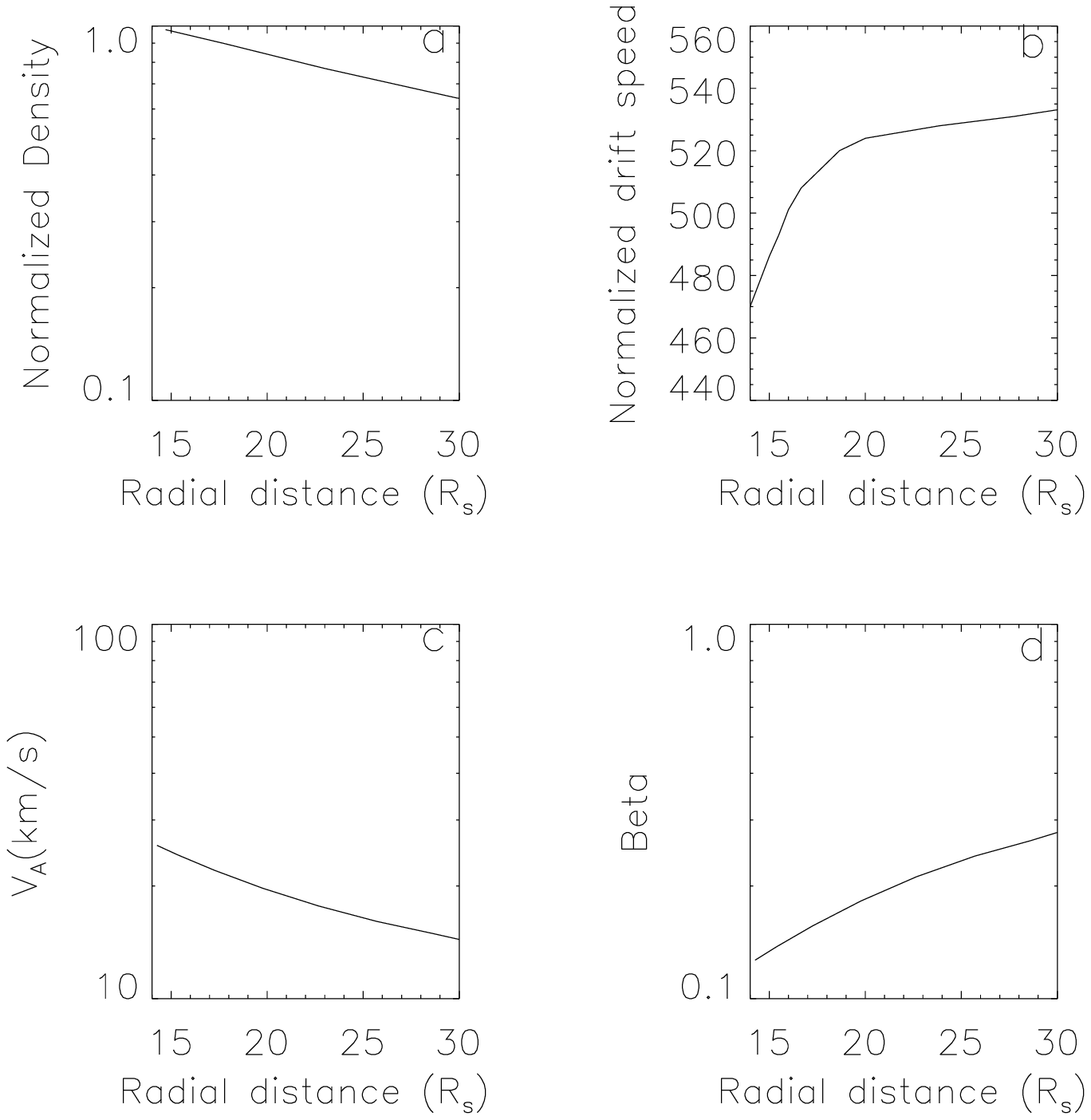} 
\end{center}
\caption{Radial profiles of (a) number density, (b) bulk velocity, (c) the Alfv\'en velocity,
and (d) the plasma beta function. Panels (a) and (b) are obtained by integrating
the proton VDF weighted by 1 and $V_z$,
respectively. Kinetic Alfv\'{e}n turbulence has been taken into account in
the evolution equation assuming a displaced Maxwellian as boundary condition
at 14 $R_{\rm s}$.}
\label{mom215}
\end{figure}

Figure 6 illustrates the VDF obtained at 30 $R_{\rm s}$ assuming a displaced
Maxwel\-lian as boundary condition at 14 $R_{\rm s}$ by the evolution equation
including the Alfv\'{e}nic turbulence terms. The top panel shows the
parallel (solid line) and perpendicular (dotted line) cross section
of the proton VDF. It can be seen that an extended nonthermal tail
is created in the direction parallel to the interplanetary magnetic field.
The Coulomb collisional diffusion appeared to be efficient enough to smooth
out a strong gradient at the turbulent diffusion boundary $V_{z}\approx
V_{\rm A}$ and to prevent formation of the quasilinear plateau at $%
V_{z}\gtrsim V_{\rm A}$. The core distribution seems not to be affected. The
bottom panel illustrates the isocontours of the proton VDF in the velocity
plane centered on the bulk velocity.

\begin{figure}[tbp]
\begin{center}
\includegraphics[width=11.5cm]{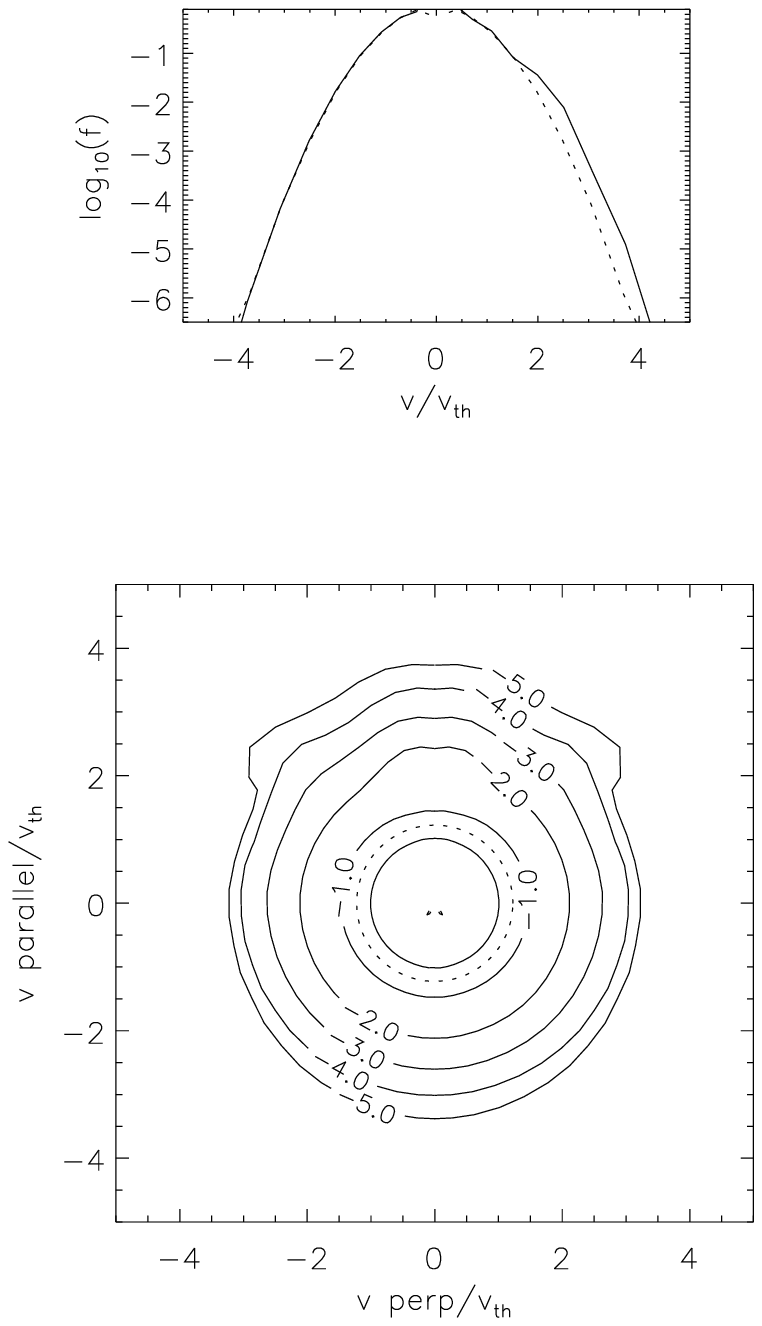} 
\end{center}
\caption{VDF obtained at 30 $R_{\rm s}$ assuming a displaced Maxwellian as the
boundary condition at 14 $R_{\rm s}$ by solving the evolution equation including the
Fokker-Planck diffusion term due to kinetic Alfv\'{e}n turbulence. Top
panel: Parallel (solid line) and perpendicular (dotted line) cross
section of the proton VDF. Bottom panel: Isocontours of the proton VDF in
the velocity plane centered on the bulk velocity.}
\label{bc2f}
\end{figure}

Particles VDF measured in-situ in the solar wind show many non-thermal
features, like particle beams, thermal anisotropies, heat fluxes, etc.,
which are possible to explain with kinetic modeling. In particular, contour
plots of the proton VDF measured around 0.3 AU by Helios showed the
nonthermal tails and sometimes also separate bumps on tails (proton beams)
formed at low radial distances \cite{Mars82, Mars2006}.

Our simulation results demonstrate that the kinetic-scale Alfv\'{e}nic
turbulence can create the nonthermal tails in the proton VDFs in the
direction of the local magnetic field. The temperature anisotropy of the
core is not obtained with this term, but could be by including the 2D
diffusion term associated to KAW.

\section{Conclusions}

An essential difference between MHD and kinetic modeling is the ability of
the latter to capture processes of the collisionless wave-particle
interactions. Kinetic effects are of particular importance in weakly and
mildly collisional media, like solar and stellar coronae and winds,
planetary magnetospheres, and interstellar medium. In particular,
permanently observed kinetic-scale Alfv\'{e}nic turbulence need to be
incorporated in kinetic solar wind modeling.

We investigated numerically the influence of the kinetic-scale Alfv\'{e}nic
turbulence on the evolution of the proton distribution function and the
principal moments in the SW at low radial distances. The effects of
the Coulomb collisions and external forces are compared to those included
also turbulence. Observed properties of Alfv\'{e}nic turbulence are taken
into account, including spectral breaks, variable spectral slopes, and
radial dependence of the turbulence level. We found significantly different
proton VDFs with and without turbulence. In particular, the obtained
simulation results demonstrate that the Alfv\'{e}nic turbulence with
characteristics typical for the solar wind is very efficient in producing
the nonthermal tails in the proton VDFs in the direction of the background
magnetic field. We observed in our simulations a fast generation of
the nonthermal proton tails at the heliocentric distances less than 0.3 AU.
The tails do not appear without turbulence terms. This provides a natural
explanation for the proton VDF with tails routinely observed in situ at
0.3-1 AU.

The turbulence effects on the number density and bulk velocity are also
investigated. It is shown that the turbulence has a non negligible influence
on the bulk velocity and increases it in the formation region of the proton
tails. On the contrary, the particle number density decreases when the
turbulence is accounted for.

In this first kinetic simulation we considered a simplest case of the
turbulence spectrum with a single spectral kink, which did not allow us to
study the influence of steeper spectra with indices $\delta _{\perp }\propto
3-4$ often observed in the solar wind. Again, we used a simplest 1D version
of the diffusion coefficient reduced by the averaging in the cross-field
velocity plane.

The bump-in-the tail VDFs of protons (proton beams), observed in the solar
wind, are not easy to reproduce by the stationary Fokker-Planck
diffusion. As is explained in our accompanying paper (Voitenko and
Pierrard, 2012), a time-varying diffusion can lead to the beam formation by
the time-of-flight effects. The beams can be also produced by the proton
trapping and acceleration in the parallel electric potential of the
intermittent KAW pulses, as is explained by Pierrard and Voitenko (2010)
and demonstrated numerically by Li {\it et al.} (2010). Simulation of
this process requires a more complex modeling of the solar wind turbulence
including intermittence. This work is in progress.

\begin{acks}
The research leading to these results has received funding from the Belgian Federal Science Policy in the framework of the program Interuniversitary Attraction Pole for the project P7/08 CHARM and from the
European Commission's FP7 Program for the STORM (313038) and SWIFF projects (263340, swiff.eu). We thank the referee for his useful suggestions.
\end{acks}

\end{article}
\end{document}